\documentclass[aps,prl,twocolumn,showpacs]{revtex4}
\usepackage{amsfonts}
\usepackage{natbib}
\usepackage{amsmath}
\usepackage{amsthm}
\usepackage{graphics}
\usepackage{subfigure}
\usepackage{amssymb}
\usepackage{graphicx}
\usepackage{color}
\usepackage{natbib}

\def\kmax{k_{\rm max}}
\newcommand\bet{{g}}
 
\newcommand\alps{{\frac{\hbar^2}{2m}}}
\newcommand\mub{{\mu}}

\newcommand\dertt[1]{ \frac{\partial{ #1}}{\partial t} }
\newcommand\gd{\mbox{${\bf \nabla}^{2}$}}
\newcommand\grad{\mbox{${\bf \nabla}$}}
\newcommand\psib{\overline{\psi}}

\begin{document}
\title{Anomalous vortex ring velocities induced by thermally-excited Kelvin waves and counterflow effects in superfluids}
\author{Giorgio Krstulovic}
\affiliation{Laboratoire de Physique Statistique de l'Ecole Normale 
Sup{\'e}rieure, \\
associ{\'e} au CNRS et aux Universit{\'e}s Paris VI et VII,
24 Rue Lhomond, 75231 Paris, France}
\author{Marc Brachet}
\affiliation{Laboratoire de Physique Statistique de l'Ecole Normale 
Sup{\'e}rieure, \\
associ{\'e} au CNRS et aux Universit{\'e}s Paris VI et VII,
24 Rue Lhomond, 75231 Paris, France}
\date{\today}
\begin{abstract}
Dynamical counterflow effects on vortex evolution under the truncated Gross-Pitaevskii equation are investigated. Standard longitudinal mutual friction effects are produced and a dilatation of vortex rings is obtained at large counterflow.
A strong temperature-dependent anomalous slowdown of vortex rings is observed and attributed to the presence of thermally exited Kelvin waves. This generic effect of finite-temperature superfluids is estimated using energy equipartition and
orders of magnitude are given for weakly interacting Bose-Einstein condensates and superfluid $^4{\rm He}$.
The relevance of thermally excited Kelvin waves is discussed in the context of quantum turbulence.
\end{abstract}
\maketitle
Quantum vortices present in superfluids interact with the normal fluid producing \emph{mutual friction} effects that must be phenomenologically introduced into Landau's two-fluid model \cite{Landau6Course,Vinenxxx}. For superfluid $^4{\rm He}$, there is no generally-accepted theory of mutual friction that is valid over the entire temperature range \cite{Donne}.
For Bose-Einstein Condensates (BEC), the Gross-Pitaevskii equation (GPE) is a dynamical description that was thought to be valid only in the low-temperature limit \cite{Proukakis:2008p1821}. 
Davis et al. \cite{Davis:2001p1475} suggested that, when a truncation of Fourier modes is performed, the resulting truncated GPE (TGPE) can also describe the (classical) thermodynamic equilibrium of homogeneous BEC \cite{Davis:2001p1475}. The TGPE was found to relax toward (microcanonical) equilibrium and a condensation transition was obtained \cite{Davis:2001p1475,Connaughton:2005p1744}. Vortex dynamics was studied within the TGPE by Berloff and Youd \cite{Berloff:2007p423} who observed a dissipative contraction of vortex rings.

The purpose of this Letter is to investigate mutual friction and counterflow effects in the context of the TGPE.
We present a stochastic algorithm that allows to efficiently generate grand canonical equilibrium states with non-zero momentum at given (target) values of temperature chemical potential and counterflow. These states are then combined with lattices of straight vortices and vortex rings and their TGPE evolutions are monitored.  
Our main result is that, beside the phenomenologically expected counterflow effects, the TGPE also induces a (phenomenologically) unexpected slowdown of vortex rings that is caused by thermally excited Kelvin waves and should be considered in quantum turbulence.

The TGPE describing a homogeneous BEC of volume $V$ is obtained from the GPE by truncating the Fourier transform of the wavefunction $\psi$: $\hat{\psi}_{\bf k}\equiv0$ for $|{\bf k}|>\kmax$ \cite{Davis:2001p1475,Proukakis:2008p1821}.
Introducing the Galerkin projector $\mathcal{P}_{\rm G}$ that reads in Fourier space $\mathcal{P}_{\rm G} [ \hat{\psi}_{\bf k}]=\theta(\kmax-|{\bf k}|)\hat{\psi}_{\bf k}$ with $\theta(\cdot)$ the Heaviside function, the TGPE  explicitly reads
\begin{equation}
i\hbar\dertt{\psi}  =\mathcal{P}_{\rm G} [- \alps \gd \psi + \bet\mathcal{P}_{\rm G} [|\psi|^2]\psi ],
\label{Eq:TGPEphys}
\end{equation}
where $|\psi|^2$ is the number of condensed particles per unit volume, $m$ is their mass and $g=4 \pi  \tilde{a} \hbar^2 / m$, with $\tilde{a}$ the $s$-wave scattering length. The superfluid velocity reads ${\bf v_{\rm s}}=(\hbar/m){\bf \nabla}\phi$, where $\phi$ is the phase of the (complex) $\psi$ and $h/m$ is the Onsager-Feynman quantum of velocity circulation around vortex lines $\psi=0$  \cite{Proukakis:2008p1821}.
When Eq. \eqref{Eq:TGPEphys} is linearized around a constant  $\psi= \hat{\psi}_{\bf 0}$, the sound velocity is given by $c={(g| \hat{\psi}_{\bf 0}|^2/m)}^{1/2}$ with dispersive effects taking place for length scales smaller than the coherence length $\xi={(\hbar^2/2m|\hat{\psi}_{\bf 0}|^2g) }^{1/2}$ that also corresponds to the vortex core size.

Equation (\ref{Eq:TGPEphys}) exactly conserves the energy $H=\int d^3 x\left( \alps |\grad \psi |^2 +\frac{g}{2}[\mathcal{P}_{\rm G}|\psi|^2]^2 \right)$ and the number of particles $N=\int  d^3 x|\psi|^2$. The momentum ${\bf P}=\frac{i\hbar}{2}\int d^3x\left( \psi {\bf \nabla}\psib - \psib {\bf \nabla}\psi\right)$ is also conserved when standard Fourier pseudo-spectral methods are used, provided that they are dealiased using the $2/3$-rule ($\kmax=2/3\times M/2$ \cite{Got-Ors} at resolution $M$) \footnote{Global momentum conservation is mandatory to correctly describe vortex-normal fluid interactions. When the nonlinear term in Eq.\eqref{Eq:TGPEphys} is written, as in  \cite{Davis:2001p1475}, $\mathcal{P}_{\rm G} [|\psi|^2 \psi]$ dealiasing must be performed at  $\kmax=M/4$.}.
%

Microcanonical equilibrium states are known to result from long-time integration of TGPE \cite{Davis:2001p1475,Connaughton:2005p1744,Berloff:2007p423}. Grand canonical states are given by the probability distribution $\mathbb{P}_{\rm st}[\psi]=\mathcal{Z}^{-1}\exp[{-\beta (H-\mu N -{\bf v_{\rm n}}\cdot {\bf P})}]$. They allow to directly control the temperature (instead of the energy in a microcanonical framework). These states can be efficiently obtained  by constructing a stochastic process that converges to a realization with the probability $\mathbb{P}_{\rm st}[\psi]$ \cite{L1-Krstulovic:PhdThesis}. This process is defined by a Langevin equation consisting in a stochastic Ginbzurg-Landau equation (SGLE):
\begin{eqnarray}
\nonumber\hbar\dertt{\psi} &=&\mathcal{P}_{\rm G}  \left[\alps \gd \psi - \bet\mathcal{P}_{\rm G} [|\psi|^2]\psi\right] \\
&&+ \mathcal{P}_{\rm G} \left[\mub \psi-i\hbar{\bf v_{\rm n}}\cdot{\bf\nabla} \psi\right]  +\sqrt{\frac{2 \hbar}{V\beta }}  \mathcal{P}_{\rm G} \left[\zeta({\bf x},t)\right] \label{Eq:SGLRphys},\hspace{3mm}
\end{eqnarray}
where the white noise  $\zeta({\bf x},t)$ satisfies $\langle\zeta({\bf x},t)\zeta^*({\bf x'},t')\rangle=\delta(t-t') \delta({\bf x}-{\bf x'})$, $\beta$ is the inverse temperature, $\mu$ the chemical potential and ${\bf v_{\rm n}}$ the normal velocity. 
The term $i\hbar\,{\bf v_{\rm n}}\cdot{\bf\nabla} \psi$  induces an asymmetry in the repartition of sound waves and generates non-zero momentum states. These states do not generally correspond to a condensate moving at velocity ${\bf v_{\rm s}}={\bf v_{\rm n}}$ because ${\bf v_{\rm s}}$ is the gradient of a phase and takes discrete values for finite size systems.
Equilibrium states states with nonzero values of the counterflow ${\bf w}={\bf v_{\rm n}-v_{\rm s}}$ are generated in this way. 

Using this algorithm in \cite{L1-Krstulovic:PhdThesis} the microcanonical and grand canonical ensembles were shown to be equivalent and the condensation transition reported in \cite{Davis:2001p1475,Connaughton:2005p1744} identified with the standard second order $\lambda$-transition. 
All the SGLE equilibrium used in this letter have a condensate at rest (${\bf v_{\rm s}=0}$) and therefore ${\bf v_{\rm n}}={\bf w}$.

At low-temperature the partition function $\mathcal{Z}$ can be exactly computed by the steepest-descent method \cite{L1-Krstulovic:PhdThesis}. In particular, setting  ${\bf v_{\rm n}}=(0,0, v_{\rm n})$ the momentum and the number of particles of the equilibrium state read $\overline{ P_z}= \frac{ \mathcal{N}}{\beta}   \frac{  m}{\mu}  f\left[\frac{4 m \mu
   }{\hbar^2k_{\rm max}^2}\right] v_{\rm n}$ and $\overline{N}=\frac{V\mu}{g}-\frac{\mathcal{N}}{\beta}\frac{3}{2\mu} f\left[\frac{4 m \mu
   }{\hbar^2k_{\rm max}^2}\right]$,
 where $\mathcal{N}=k_{\rm max}^3 V/6 \pi ^2$ is the total number of modes and $f[z]=z-z^{3/2} \cot ^{-1}\left(\sqrt{z}\right)$.
These relations furnish an explicit expression for the normal density
$\rho_{\rm n}=\frac{1}{V}\left.\frac{\partial P_{z}}{\partial {v_{\rm n}}}\right|_{{v_{\rm n}}=0}$.

\begin{figure}[htbp]
\begin{center}
\includegraphics[height=2.8cm]{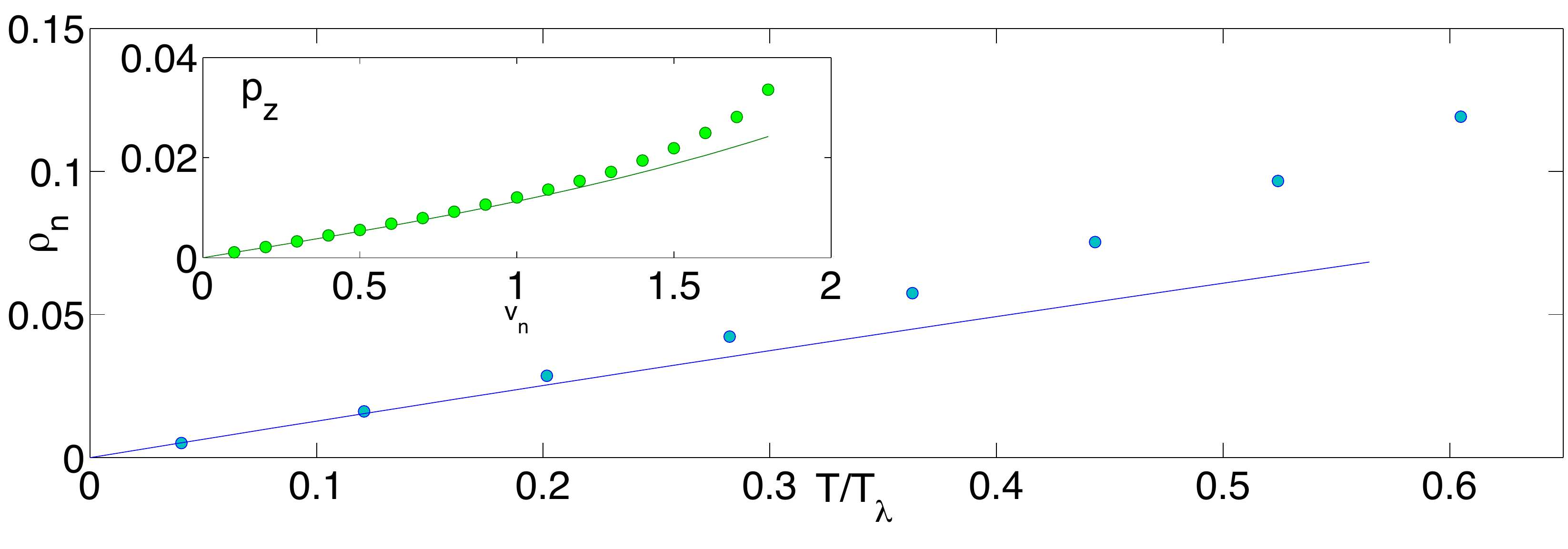}
\caption{Temperature dependence of the normal density $\rho_{\rm n}$ (see text). Inset: $P_{z}$ as a function of $v_{\rm n}$ at fixed temperature $T=0.08T_\lambda$. Points: SGLE \eqref{Eq:SGLRphys} equilibration at resolution $64^3$; solid lines: low-temperature exact results.}
\label{Fig:rhon}
\end{center}
\end{figure}

The direct control of the counterflow $v_{\rm n}$ in the SGLE algorithm allows to obtain the temperature dependence of $\rho_{\rm n}$ in the TGPE context.  
Low-temperature exact results are in good agreement with SGLE data, see Fig.\ref{Fig:rhon}.

In all the numerical simulations presented in this Letter $\mu$ is adjusted in order to fix the density $\rho=m N/V$ to $1$ and the physical constants in Eqs.\eqref{Eq:TGPEphys} and \eqref{Eq:SGLRphys} are determined by the relations $\xi \kmax=1.48$ and  $c=2$. The inverse temperature is normalized as $\beta=\mathcal{N}/VT$ and $V=(2\pi)^3$. With these choice of parametrization the $\lambda$-transition temperature is independent of $\mathcal{N}$ and its value is fixed to $T_\lambda=2.48$; the quantum of circulation $h/m$ has the value $c\,\xi/\sqrt{2}$. 

\begin{figure}[htbp]
\begin{center}
\includegraphics[height=7cm]{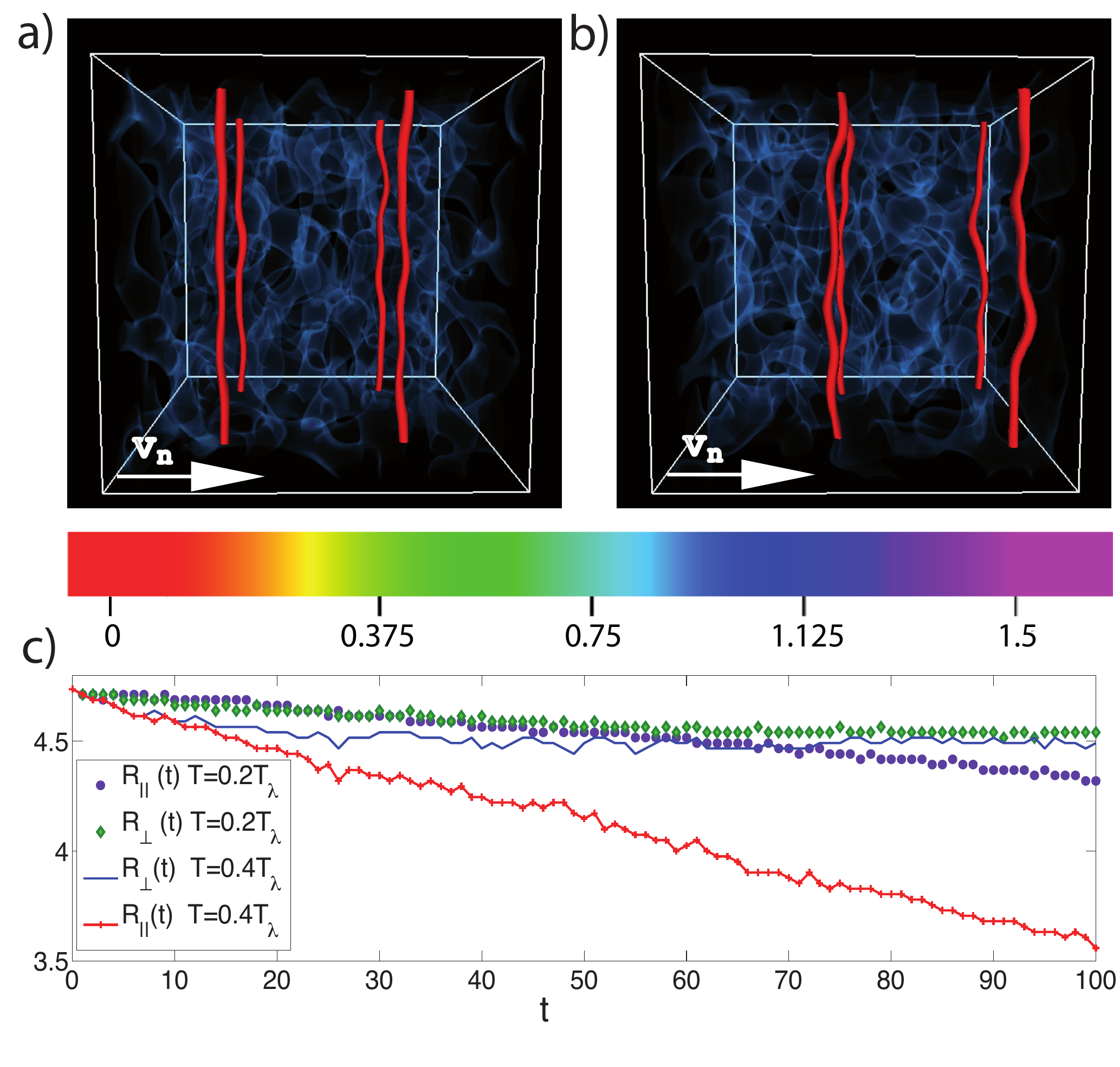}
\caption{a-b) Density at $t=0$, $100$ of the lattice configuration (red) with $T= 0.4\,T_\lambda$ and $v_{\rm n}=0.4$. Blue clouds correspond to density fluctuations. c) Positions $(R_{\parallel},R_{\perp})$ of a single vortex for $T=0.2\,T_\lambda$, $ T=0.4\,T_\lambda$ and $v_{\rm n}=.4$. Resolution $64^3$.}
\label{Fig:Lattice}
\end{center}
\end{figure}

We now turn to counterflow effects. 
To wit, we use an array of alternate-sign straight vortices $\psi_{\rm lattice}$ (see \cite{Nore:1994p405}).
This exact stationary solution of the GPE is obtained by a Newton method. 
The vortices are separated by a distance $\pi$ and can be considered isolated when $\xi\to0$, as the resolution is increased.
An equilibrium state $\psi_{\rm eq}$ is prepared using the SGLE \eqref{Eq:SGLRphys} with counterflow $v_{\rm n}$ perpendicular to the vortices. The initial condition $\psi=\psi_{\rm lattice}\times\psi_{\rm eq}$ is then evolved with the TGPE. Figure \ref{Fig:Lattice}.a displays $3$D visualizations of the density at $t=0$ and $t=100$ where the displacement of the lattice is apparent. 
The temporal evolution of the (parallel and perpendicular to ${\bf v}_{\rm n}$) position of a vortex $(R_{\parallel},R_{\perp})$  
are presented on Fig.\ref{Fig:Lattice}.c for $T=0.2\,T_\lambda$, $T= 0.4\,T_\lambda$ and $v_{\rm n}=0.4$. 
%
The counterflow-induced vortex velocity clearly depends on the temperature. 
A perpendicular motion is also induced at short times. 
This motion has two phases: 
first an adaptation, making the lattice slightly imperfect,
followed by a much slower  perpendicular motion.
%
%
Observe that the imperfection of the lattice at final configurations is almost equal for the two temperatures presented in Fig.\ref{Fig:Lattice}.c, but the parallel velocities are considerably different. 
The self-induced parallel velocity caused by the slight lattice imperfection is thus very small and not driving the longitudinal motion.

\begin{figure}[htbp]
\begin{center}
\includegraphics[height=3.5cm]{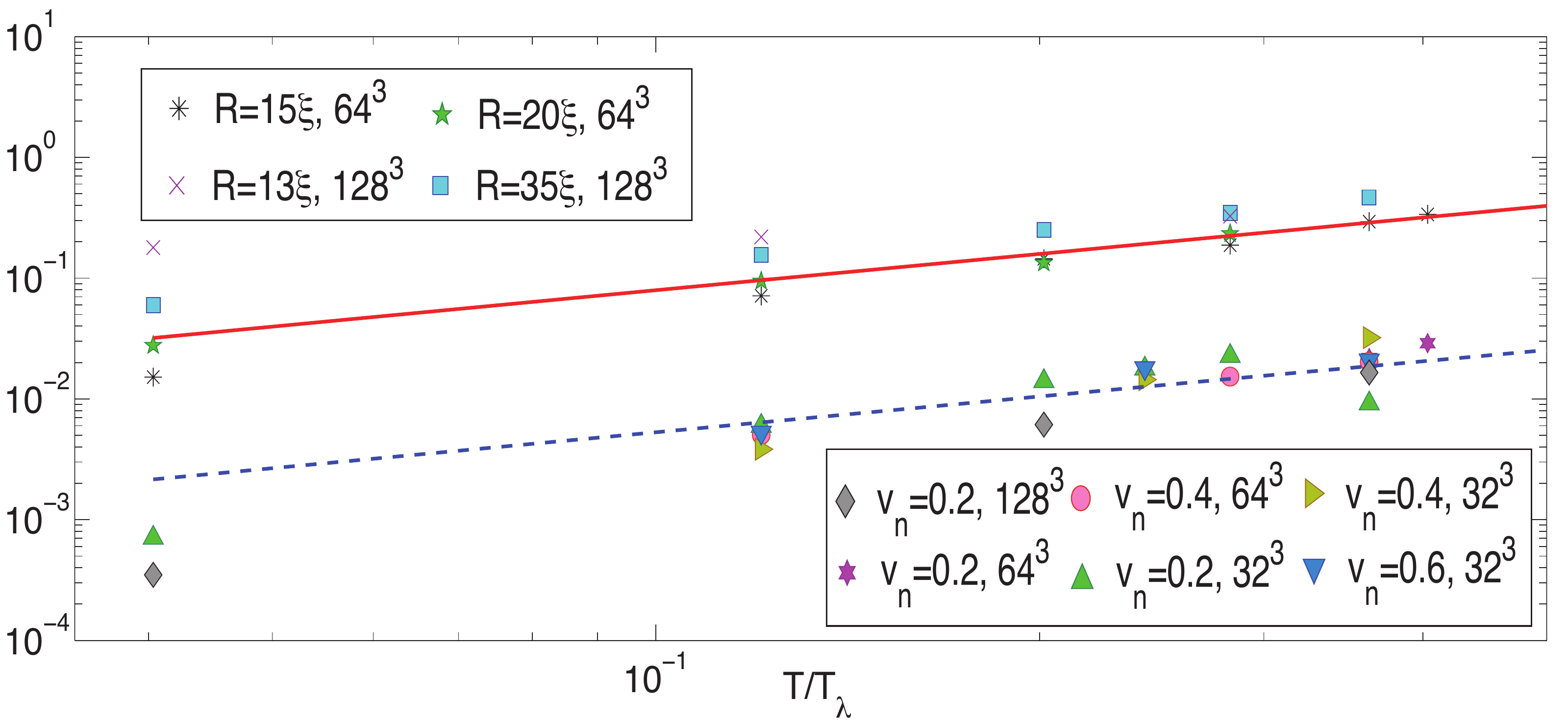}
\caption{Temperature dependence of counterflow-induced lattice velocity $v_{\parallel}/v_{\rm n}$ (bottom) and ring slowdown $\Delta v_{L}/u_{\rm i}$ (top) obtained with $v_{\rm n}=0$. Dashed line: prediction of Eq.\eqref{Eq:VortexDyn} with $\alpha'=0.83\rho_{\rm n}/2\rho$; solid line: prediction of anomalous slowdown by Eq.\eqref{Eq:AnEffect} with $R=20\xi$ at various resolutions.}
\label{Fig:alphap}
\end{center}
\end{figure}

We now concentrate on the measurement of $R_{\parallel}$ for which the present configuration is best suited.
$R_{\parallel}$ has a linear behavior, that allows to directly measure the parallel velocity $v_{\parallel}$. The temperature dependence of $v_{\parallel}/v_{\rm n}$ is presented on Fig.\ref{Fig:alphap} for different values of $v_{\rm n}$ and $\xi$. This behavior is consistent with the standard phenomenological model for the vortex line velocity ${\bf v}_{\rm L}$ \cite{Donne}:
\begin{equation} 
{\bf v}_{\rm L}={\bf v}_{\rm sl}+\alpha {\bf s'}\times({\bf v}_{\rm n}-{\bf v}_{\rm sl})-\alpha'{\bf s'}\times[{\bf s}'\times({\bf v}_{\rm n}-{\bf v}_{\rm sl})],\label{Eq:VortexDyn}
\end{equation}
where $s'$ is the tangent of the vortex line, ${\bf v}_{\rm sl}={\bf v}_{\rm s}+{\bf u}_{\rm i}$ is the local superfluid velocity with  ${\bf u}_{\rm i}$  the self-induced vortex velocity and ${\bf v_{\rm n}}$ the normal velocity. The mutual friction coefficients in Eq.\eqref{Eq:VortexDyn} are typically written as $\alpha=B\rho_{\rm n}/2\rho, \alpha'=B'\rho_{\rm n}/2\rho$ where $B$ and $B'$ are order-one and weakly temperature-dependent. Equation \eqref{Eq:VortexDyn} applied to a straight vortex with $v_{\rm n}$ perpendicular to the vortex and $v_{\rm s}=0$ yields $\alpha'=v_{\parallel}/v_{\rm n}$. 
The value of $\alpha'=B'\rho_n/2\rho$ with $B'=0.83$ is displayed on Fig.\ref{Fig:alphap} (bottom dashed line) and is in good agreement with the lattice data. 


We now turn to the interaction of vortex rings and counterflow. 
The Biot-Savart self-induced velocity of a perfectly circular vortex ring of radius $R$ is given by
\begin{equation}
u_{\rm i}=\frac{\hbar}{2m}\frac{C(R/\xi)}{R}\,,\hspace{5mm} C(z)=\ln{(8z)}-a \label{Eq:ui}
\end{equation} 
where $a$ is a core model-depending constant \cite{Donne}. We have checked, using an initial data $\psi_{\rm ring}$ prepared by a Newton method that the GPE (large $R/\xi$) ring translational velocity is well reproduced by \eqref{Eq:ui} with $a=0.615$.

\begin{figure}[htbp]
\begin{center}
\includegraphics[height=6cm]{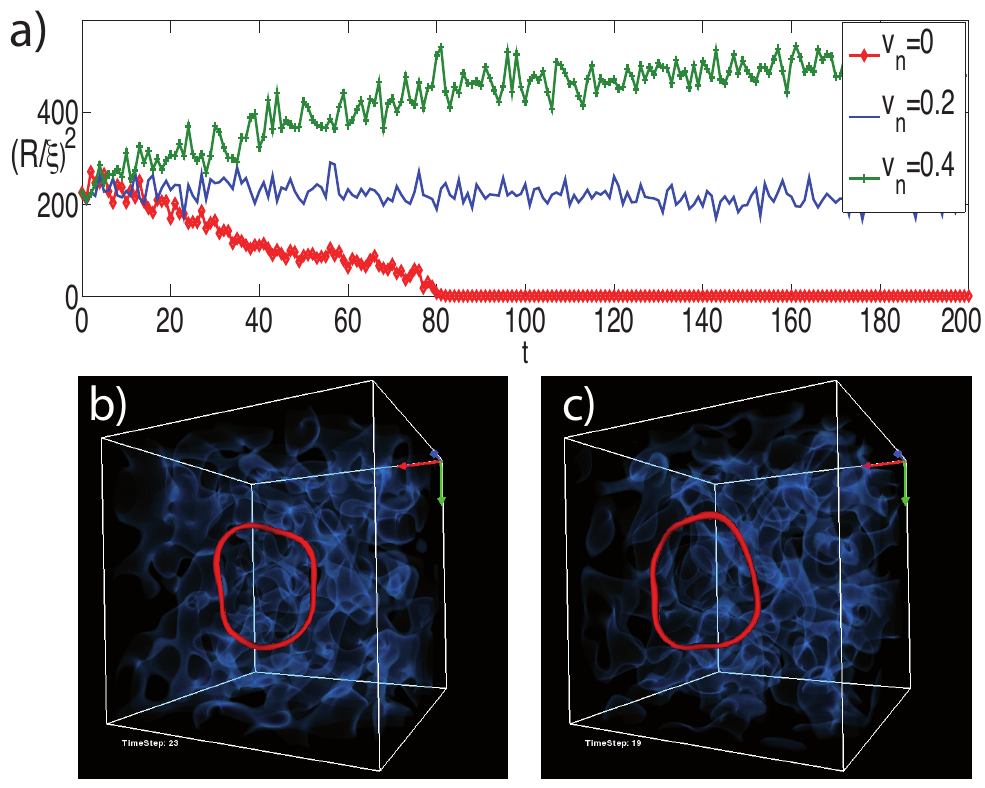}
\caption{a) Temporal evolution of the (squared) length of a vortex ring at different values of counterflow $v_{\rm n}$ (temperature $T=0.4\,T_{\lambda}$ and initial radius $R=15\xi$). b-c) $3$D visualization of vortex ring ($R=20\xi$) and density fluctuations at $t=18$, $19$, with $T=0.4\,T_{\lambda}$ and resolution $64^3$. Same colorbar as in Fig.\ref{Fig:Lattice}. Thermally-excited Kelvin waves are apparent.}
\label{Fig:3}
\end{center}
\end{figure}

Equation \eqref{Eq:VortexDyn} with $v_{\rm n}$ perpendicular to the ring and $v_{\rm s}=0$ yields the radial velocity $\dot{R}=-\alpha(u_{\rm i}-v_{\rm n})$. 
The case without counterflow ($v_{\rm n}=0$) was studied by Berloff and Youd \cite{Berloff:2007p423} and a contraction of vortex rings compatible with \eqref{Eq:VortexDyn} was reported. To study the influence of counterflow we prepare an initial condition $\psi=\psi_{\rm ring}\times\psi_{\rm eq}$ in the same way as above for the vortex lattice. The temporal evolution of the (squared) vortex length of a ring of initial radius $R=15\xi$ at temperature $T=0.4\,T_{\lambda}$ and $v_{\rm n}=0$, $0.2$ and $0.4$ is displayed on Fig.\ref{Fig:3}.a. 
The Berloff-Youd contraction \cite{Berloff:2007p423} is apparent in absence of counterflow (bottom curve). The temperature dependence of the contraction, related to the $\alpha$ coefficient in Eq.\eqref{Eq:VortexDyn}, also quantitatively agrees with their published results (data not shown).

A dilatation of vortex rings is obtained (top curve on Fig.\ref{Fig:3}.a). when the counterflow  $v_{\rm n}$ is large enough.
Such a dilatation --a hallmark of counterflow effects-- is expected  \cite{Donne} to correspond to a change of sign of ${\bf v}_{\rm n}-{\bf v}_{\rm sl}$ in Eq. \eqref{Eq:VortexDyn}. 
However, the predictions of Eq.\eqref{Eq:VortexDyn} unexpectedly turn out to be quantitatively wrong. Indeed, 
using Eq.\eqref{Eq:ui} in the conditions of Fig.\ref{Fig:3}.a one finds ${\bf v}_{\rm sl}={\bf u}_{\rm i}=0.39$
which is significantly larger than normal velocity $v_{\rm n}=0.2$ around which
dilatation starts to take place (see middle curve on Fig.\ref{Fig:3}.a).
Equation \eqref{Eq:VortexDyn} prediction for the longitudinal velocity $v_{L}=(1-\alpha')u_{\rm i}+\alpha'v_{\rm n}$ is also unexpectedly wrong. 
Using the value of $\alpha'$ determined above on the vortex array, one finds $v_{L}\sim 0.98 u_{\rm i}$ and from Eq.\eqref{Eq:ui} one finds for $v_{L}$ the value $0.38$ that is larger than the measured value $v_L=0.23$.

This anomaly of the ring velocity $v_L$ is also present in the absence of counterflow ($v_{\rm n}=0$) where Eq.\eqref{Eq:VortexDyn} predicts that $\alpha'$ should be given by $\Delta v_{L}/u_{\rm i}\equiv(u_{\rm i}-v_{L})/u_{\rm i}$. 
The temperature dependence of $\Delta v_{L}/u_{\rm i}$ is displayed on Fig.\ref{Fig:alphap} (top curve). 
Observe that $\Delta v_{L}/u_{\rm i}$ is one order of magnitude above the transverse mutual friction coefficient $\alpha'$ measured on the lattice. 

We now relate the thermally-induced anomaly to the velocity $v_{\rm a}$ induced on a vortex ring by a single a Kelvin wave of (small) amplitude $A$ and (large) wavenumber $N_{\rm K}/2\pi R$ obtained in the LIA \cite{Kiknadze:2002p3905} and Biot-Savart \cite{Barenghi:2006p3901} frameworks.
The velocity $v_{\rm a}$ reads (see Eq. (26) of \cite{Kiknadze:2002p3905})
\begin{equation}
v_{\rm a}= u_{\rm i}(1-A^2N_{\rm K}^2/R^2+3A^2/4R^2)\label{Eq:AneffKik}
\end{equation}
where  $u_{\rm i}$ is the (undisturbed) ring velocity \eqref{Eq:ui}.

The TGPE model naturally include thermal fluctuations that excite Kelvin waves as apparent on Fig.\ref{Fig:3}.b-c. 
We assume that the slowing down effect of each individual Kelvin wave is additive and that the waves populate all the possible modes. 
Kelvin waves being bending oscillations of the the quantized vortex lines their wavenumber must satisfy $k \le k_\xi= 2 \pi / \xi$. The total number of thermally excited Kelvin waves is thus
$\mathcal{N}_{\rm Kelvin}\approx R\,k_\xi$.

The amplitude term $A^2N_{\rm K}^2/R^2$ in \eqref{Eq:AneffKik} can be obtained by simple equipartition arguments. The energy of a (perfect) ring is $E=\frac{2 \pi^2\rho_{\rm s}\hbar^2}{m^2}R[C(R/\xi)-1 ]$, with $\rho_{\rm s}$ the superfluid density \cite{Donne}.  A Kelvin wave produces a variation of the ring length $\Delta L=\pi A^2 N_{\rm K}^2/R$. Its energy can thus be estimated as
$\Delta E=\frac{dE}{dR}\frac{\Delta L}{2\pi}.$
Assuming $\Delta E=\beta^{-1}$ yields, at low temperature where $\rho_{\rm s}\approx\rho$, $A^2N_{\rm K}^2/R^2=m^2\beta^{-1}/\pi^2\rho\hbar^2R\,C(R/\xi)$ \footnote{This formula predicts a $UV$-convergent r.m.s amplitude that is in good agreement with TGPE data, with values small enough to avoid self-reconnections of the ring.}.
%
%
Replacing $A^2/R^2$ in Eq.\eqref{Eq:AneffKik}, the dominant effect is obtained by summing up to $\mathcal{N}_{\rm Kelvin}$ and it finally reads:
\begin{equation}
\frac{\Delta v_{L}}{u_{\rm i}}\equiv\frac{u_{\rm i}-v_{\rm a}}{u_{\rm i}}\approx\frac{\beta^{-1}m^{2}}{\pi^2 \rho \hbar^{2}C(R/\xi)}k_\xi\label{Eq:AnEffect}.
\end{equation}
The thermally-induced anomalous slowdown \eqref{Eq:AnEffect} is in good agreement with the TGPE data displayed on Fig.\ref{Fig:alphap}.

We now extend \eqref{Eq:AnEffect} in order to take into account quantum effects and estimate orders of magnitude in the physical case of BEC and superfluid $^4{\rm He}$. 
The dispersion relation of Kelvin waves 
$\omega(k)=\frac{\hbar}{2m}k^2C(R/\xi)$ \cite{Kiknadze:2002p3905}
implies (using the relation $\hbar\omega(k_{\rm eq})=\beta^{-1}=k_{\rm B}T$) that
Kelvin waves are not in equipartition for wavenumbers $k>k_{\rm eq}=(2m k_{\rm B}T/\hbar^2C(R/\xi))^{1/2}$, as (like in blackbody radiation) quantum effects are relevant in this range.

For weakly-interacting BEC with mean inter-atomic particle distance $\ell \sim |\hat{\psi_{\bf 0}}|^{-2/3}$ satisfying $\tilde{a}\ll\ell\ll \xi$ the condensation temperature is $T_\lambda\sim\hbar^2/k_{\rm B} m\ell^2$. For $T>T^*$, where $T^*/T_\lambda\sim C(R/\xi)\, \ell^2/\xi^2\ll1$, it is straightforward to show that $k_{\rm eq}>k_\xi$ and therefore that \eqref{Eq:AnEffect} directly applies and reads $\Delta v_{L}/u_{\rm i}\sim (\ell/\xi)(T/T_\lambda \,C(R/\xi))$.
%
For $T<T¬^*$, $k_\xi$ must be replaced by $k_{\rm eq}$ in formula \eqref{Eq:AnEffect} and the slowdown becomes $\Delta v_{L}/u_{\rm i}\sim(T/T_\lambda\, C(R/\xi))^{3/2}$.

At zero-temperature it is natural to suggest that the quantum fluctuations of the amplitudes of Kelvin waves produce an additional effect. 
This effect can be estimated by using $\Delta E=\hbar\omega(k)/2$. It is radius-independent and of order $\Delta v_{L}/u_{\rm i}\sim (\ell/\xi)^{3}$ (see \cite{L1-Krstulovic:PhdThesis}). It is interesting to note that a drift effect of order $(\ell/\xi)^{3}$ is also obtained by balancing with a Magnus force \cite{Donne} the Roberts and Pomeau \cite{Roberts:2005p5897} Casimir-like force due to the scattering of (density) zero-temperature quantum fluctuations.

In a low-$T$ physical BEC, with quantum distribution of sound waves,  $\rho_n/\rho\sim (T/T_\lambda)^4$ \cite{Landau6Course}  and the standard effects \eqref{Eq:VortexDyn} are of order $(T/T_\lambda)^4$. Thus the new effect should dominate in this limit.
In the case of superfluid $^4{\rm He}$ the GPE description is only expected to give qualitative predictions \cite{Donne}. 
Nevertheless the new effect should also be dominant at low-temperature. 

Thermally excited Kelvin waves 
can also be relevant in the broader context of quantum turbulence where
Kelvin waves are excited at low temperature by vortex reconnection and their energy finally decays into sound waves \cite{Vinen:2008p2925}.  
The phenomenological speed of vortex rings (Eqs.\eqref{Eq:VortexDyn} and \eqref{Eq:ui}) is routinely used in this context, clearly an overestimation as these rings are thermally perturbed and hence, in reality, slower.
In a dilute gas of vortex rings this slowdown will increase the time between collisions and inhibit reconnection by a factor given by \eqref{Eq:AnEffect}.
In the case of a dense vortex tangle their effect is more difficult to predict but should be, in the future, studied using the TGPE.

In summary we obtained and measured standard counterflow mutual friction effects
within the TGPE.
Our main result is that vortex rings are decelerated by thermal fluctuations of Kelvin waves and
that these fluctuations, generic of finite-temperature superfluids, produce an experimentally-testable effect that dominates 
the standard effects at low-temperature.

We acknowledge useful scientific discussions with C. Barenghi. 
The computations were carried out at IDRIS (CNRS) and visualizations used VAPOR \footnote{http://www.vapor.ucar.edu}.

\bibliographystyle{apsrev}


\end{document}